\documentclass[reprint,amsmath,amssymb,aps,superscriptaddress]{revtex4-2}
\usepackage{graphicx}
\usepackage{multirow}%
\usepackage{amsmath,amssymb,amsfonts}%
\usepackage{amsthm}%
\usepackage{dcolumn}
\usepackage{bm}
\usepackage[T1]{fontenc}

\begin{document}

\title{Observation of Switchable Chiral Magnons in an Altermagnet}
\author{Zheyuan Liu}
\affiliation{Institute for Solid State Physics, the University of Tokyo, Kashiwa 277-8581, Japan}
\author{Hodaka Kikuchi}
\affiliation{Institute for Solid State Physics, the University of Tokyo, Kashiwa 277-8581, Japan}
\affiliation{Neutron Scattering Division, Oak Ridge National Laboratory, Oak Ridge, TN 37831, USA}
\author{Zijun Wei}
\affiliation{Institute for Solid State Physics, the University of Tokyo, Kashiwa 277-8581, Japan}
\author{Shinichiro Asai}
\affiliation{Institute for Solid State Physics, the University of Tokyo, Kashiwa 277-8581, Japan}
\author{Mechthild Enderle}
\affiliation{Institut Laue-Langevin, Grenoble, CS 20156, France}
\author{Ursula B. Hansen}
\affiliation{Institut Laue-Langevin, Grenoble, CS 20156, France}
\author{Vasile O. Garlea}
\affiliation{Neutron Scattering Division, Oak Ridge National Laboratory, Oak Ridge, TN 37831, USA}
\author{Manh D. Le}
\affiliation{ISIS Neutron and Muon Source, Rutherford Appleton Laboratory, Didcot, OX11 0QX, UK}
\author{G\o ran J. Nilsen}
\affiliation{ISIS Neutron and Muon Source, Rutherford Appleton Laboratory, Didcot, OX11 0QX, UK}
\author{Igor A. Zaliznyak}
\affiliation{Condensed Matter Physics and Materials Science Division, Brookhaven National Laboratory, Upton, NY 11973, USA}
\author{Takatsugu Masuda}
\affiliation{Institute for Solid State Physics, the University of Tokyo, Kashiwa 277-8581, Japan}
\affiliation{Institute of Materials Structure Science, High Energy Accelerator Research Organization, Ibaraki 305-0801, Japan}
\affiliation{Trans-scale Quantum Science Institute, the University of Tokyo, Tokyo 113-0033, Japan}

\date{\today}

\begin{abstract}
Chiral magnons, the quanta of handed spin waves, transport spin angular momentum without energy loss due to Joule heating. The recently discovered altermagnets were proposed to host chiral magnons arising from a non-relativistic exchange mechanism, similar to that in ferromagnets but without net magnetization, offering a stray-field-free platform for efficient magnon spin-current manipulation. In this work, we directly observed chiral magnons in the altermagnetic prototype MnTe using polarized inelastic neutron scattering. Furthermore, the magnon chirality was found to be reversibly switched by magnetic-field control, establishing a robust foundation for functional altermagnetic magnonics. 
\end{abstract}

\maketitle

Efforts to exploit the spin degree of freedom in electronics have stimulated the utilization of magnons, the collective excitations of ordered electron spins~\cite{kajiwaraTransmissionElectricalSignals2010}. Because of their spin angular momentum and immunity to Coulomb scattering, magnons serve as a promising medium for long-distance information transport and readout~\cite{cornelissenLongdistanceTransportMagnon2015,lebrunTunableLongdistanceSpin2018,yuanExperimentalSignaturesSpin2018,liSpinCurrentSubterahertzgenerated2020}. Meanwhile, the spin torque delivered by the magnons is powerful enough to switch magnetic domains for writing operations~\cite{wangMagnetizationSwitchingMagnonmediated2019,hanMutualControlCoherent2019}. The coherence of magnons further enables nearly dissipationless ballistic transport and interference-based processing~\cite{chumakMagnonSpintronics2015,hanMutualControlCoherent2019,pirroAdvancesCoherentMagnonics2021,hortensiusCoherentSpinwaveTransport2021}. In addition, nontrivial topology in quantum magnets~\cite{yaoTopologicalSpinExcitations2018,yuanDiracMagnonsHoneycomb2020,chenMagneticFieldEffect2021} protects magnonic transport in chiral edge modes~\cite{huTunableMagnonicChern2022,zhangSwitchableLongdistancePropagation2025}, leading to exotic phenomena such as the magnon thermal Hall effect~\cite{onoseObservationMagnonHall2010} and the magnon spin Nernst effect~\cite{chengSpinNernstEffect2016,zyuzinMagnonSpinNernst2016}, thereby opening up broader prospects for magnonics.

The net magnetization in ferromagnets (or ferrimagnets) breaks time-reversal symmetry (TRS), giving rise to chiral magnons~\cite{jenniChiralityMagneticExcitations2022,nambuObservationMagnonpolarization2020} capable of carrying magnon spin-current~\cite{kajiwaraTransmissionElectricalSignals2010,guObservingDifferentialSpin2025}, a crucial enabler of magnonics. However, the presence of net magnetization fundamentally limits device compatibility and miniaturization.
In contrast, in collinear antiferromagnets, the net magnetization is eliminated and TRS is preserved in the non-relativistic limit, resulting in degenerate, mutually compensating chiral magnons~\cite{rezendeIntroductionAntiferromagneticMagnons2019}. 
Moreover, in the presence of crystal field spin anisotropy, these modes become mixed, and magnons generally lose their chirality. 
Consequently, generating magnon spin-currents in antiferromagnets typically requires either a canted moment induced by an external magnetic field~\cite{lebrunTunableLongdistanceSpin2018,yuanExperimentalSignaturesSpin2018,liSpinCurrentSubterahertzgenerated2020} or topological edge states driven by relativistic spin--orbit coupling~\cite{chengSpinNernstEffect2016,zyuzinMagnonSpinNernst2016}.

Recently, a new fundamental class of  magnets, altermagnets, has been proposed to elegantly resolve this long-standing dilemma~\cite{smejkalConventionalFerromagnetismAntiferromagnetism2022,yuanPredictionLowZCollinear2021,hayamiBottomupDesignSpinsplit2020,nakaSpinCurrentGeneration2019}: their distinct spin symmetry spontaneously breaks TRS without net magnetization, leading to the emergence of intrinsic, nonrelativistic chiral magnons  in the absence of spin-orbit coupling~\cite{liuChiralSplitMagnon2024,sunObservationChiralMagnon2025,mcclartyObservingAltermagnetismUsing2025,smejkalChiralMagnonsAltermagnetic2023,nakaSpinCurrentGeneration2019,chenUnconventionalMagnonsCollinear2025}. Owing to their exchange origin, altermagnetic chiral magnons are expected to transport spin currents as efficiently as ferromagnetic magnons~\cite{cuiEfficientSpinSeebeck2023,weissenhoferAtomisticSpinDynamics2024}, offering promising functionalities for magnonics.
Following its theoretical prediction, altermagnetism has been rapidly confirmed experimentally. Photoemission spectroscopy revealed spin-split electronic bands~\cite{krempaskyAltermagneticLiftingKramers2024,leeBrokenKramersDegeneracy2024,osumiObservationGiantBand2024,dingLargeBandSplitting2024,jiangMetallicRoomtemperatureDwave2025,zhangCrystalsymmetrypairedSpinValley2025}, and inelastic neutron scattering (INS) detected magnon band splitting~\cite{liuChiralSplitMagnon2024,sunObservationChiralMagnon2025}. The circular dichroism signal of magnon excitations measured by resonant inelastic X-ray scattering has been reported to primarily probe static magnetic domains, rather than dynamical chiral magnons~\cite{takegamiCircularDichroismResonant2025}. Thus, the most decisive hallmark of altermagnetism---the chiral magnons, i.e., their handed spin character---has remained experimentally unverified.

\begin{figure*}[htbp]
	\centering
	\includegraphics[width=\textwidth]{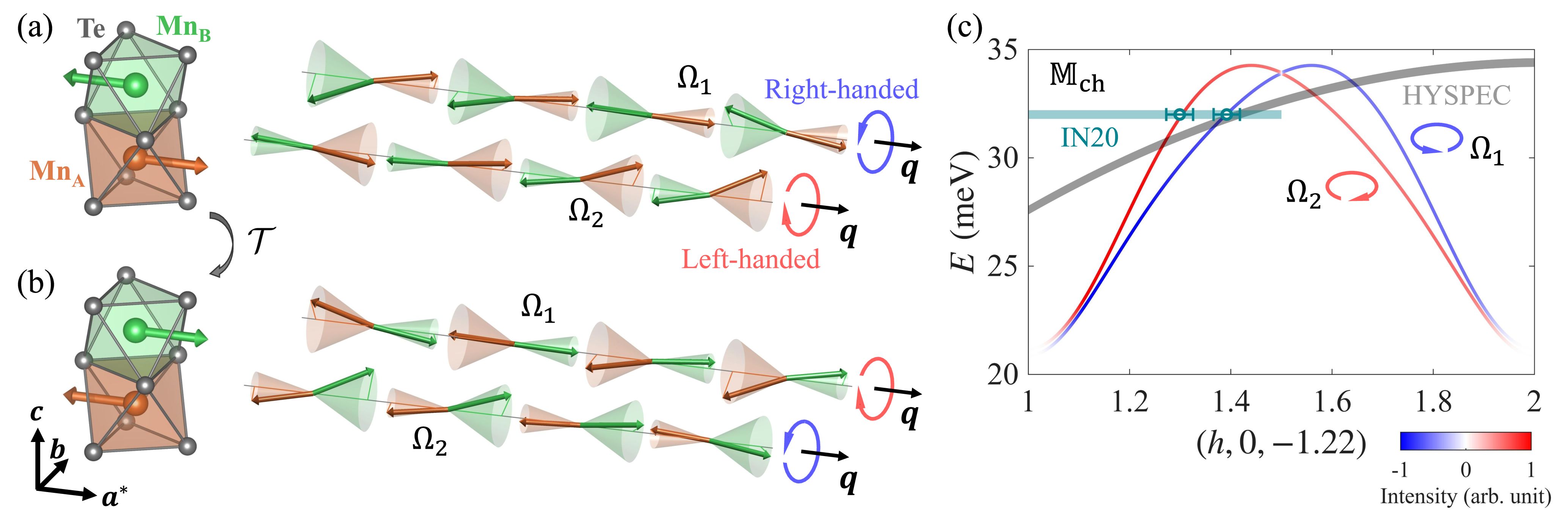}
	\caption{(a) Left: Crystal and magnetic structure of MnTe. Right: Classical schematic of chiral magnon modes in an altermagnet. The two nondegenerate modes, $\Omega_1$ and $\Omega_2$, possess right- and left-handed chirality, respectively. (b) Left: Antiphase magnetic domains corresponding to the left panel of (a). Right: The chiralities of the corresponding magnon modes, opposite to those in the right panel of (a). (c) Calculated neutron chiral term $M_\text{ch}$ for the magnon modes $\Omega_1$ and $\Omega_2$ with the domain state in (a). The gray curve indicates the $\boldsymbol{Q}\!-\!E$ coverage in the HYSPEC experiment. The blue line indicates the constant-$E$ scan in the IN20 experiment. The data points with error bars denote the observed peak centers and FWHMs of the constant-$E$ scans on IN20.
	}\label{fig1}
\end{figure*}

Here we report the direct observation of chiral magnons in an altermagnet by polarized INS (PINS). 
We study the prototypical altermagnet MnTe, which hosts two opposite-spin sublattices, Mn$_\text{A}$ and Mn$_\text{B}$, related by a sixfold improper rotation, as shown by the left panel of Fig.~\ref{fig1}(a). The magnetic easy axis lies along the crystallographic $a^{*}$ (or $b^{*}$) axis. Altermagnetic exchange lifts the degeneracy of the magnon modes $\Omega_1$ and $\Omega_2$ with opposite handedness~\cite{liuChiralSplitMagnon2024,mcclartyObservingAltermagnetismUsing2025}, as shown by the right panel of Fig.~\ref{fig1}(a). Because TRS is broken, two antiphase magnetic domains with antiparallel N{\'e}el vectors ($\boldsymbol{n}=\boldsymbol{S}_\text{A}-\boldsymbol{S}_\text{B}$, where $\boldsymbol{S}_\text{A}$ and $\boldsymbol{S}_\text{B}$ are the spin moments of Mn$_\text{A}$ and Mn$_\text{B}$) are present. The chirality of $\Omega_1$ and $\Omega_2$ reverses under $\boldsymbol{n}\to -\boldsymbol{n}$ and is therefore opposite in the two antiphase domains, as shown in the Figs.~\ref{fig1}(a) and~\ref{fig1}(b). In addition, the $C_3$ lattice symmetry yields three 120$^\circ$ magnetic orientational domains.

Consequently, the net magnon chirality is compensated unless a domain imbalance yields a finite sample-averaged net N{\'e}el vector $\overline{\hat{\boldsymbol{n}}}$ [with $\hat{\boldsymbol{n}}=\boldsymbol{n}/n$; the overbar denotes averaging over all six magnetic domains (see Supplemental Material (SM)~\cite{Methods} for detail)]. A magnetic field applied along the $c$ axis lifts the degeneracy between the antiphase domains~\cite{mazinOriginGossamerFerromagnetism2024,kluczykCoexistenceAnomalousHall2024}, enabling control of $\overline{\hat{\boldsymbol{n}}}$ by magnetic-field cooling (FC, see SM~\cite{Methods} for detail) through the magnetic ordering temperature $T_\text{N}=307$ K~\cite{harikiXRayMagneticCircular2024,aminNanoscaleImagingControl2024}. Furthermore, in the MnTe crystals studied here the orientational-domain populations are naturally imbalanced, likely due to inhomogeneous strain, thereby producing a finite $\overline{\hat{\boldsymbol{n}}}$ after FC with the field along the $c$ axis.

Polarized neutron scattering experiments were performed on two MnTe single-crystal samples, S1 and S2. Sample S1 was measured on the time-of-flight spectrometer HYSPEC~\cite{zaliznyakPolarizedNeutronScattering2017}, and sample S2 on the triple-axis spectrometer IN20~\cite{kuldaIN20ILLHighflux2002}, with the scattering plane set to $(h0l)$ in both cases. We adopt the Blume-Maleyev coordinate system for the scattering geometry, where $\boldsymbol{x}$ is parallel to the momentum transfer $\boldsymbol{Q}$, $\boldsymbol{y}$ is perpendicular to $\boldsymbol{x}$ and lies in the scattering plane, and $\boldsymbol{z}=\boldsymbol{x}\times\boldsymbol{y}$ (with $\boldsymbol{z}$ parallel to the $b$ axis, i.e., $(-k,2k,0)$). The measured neutron intensity for a given spin-polarization condition is denoted $I_\alpha^{if}$, where $\alpha=x,y,z$ specifies the polarization axis and $i,f$ indicate the spin-up ($+$), spin-down ($-$), or not analyzed ($0$) states of the incident and scattered neutrons, respectively. 

\begin{figure*}[t]
	\centering
	\includegraphics[width=\textwidth]{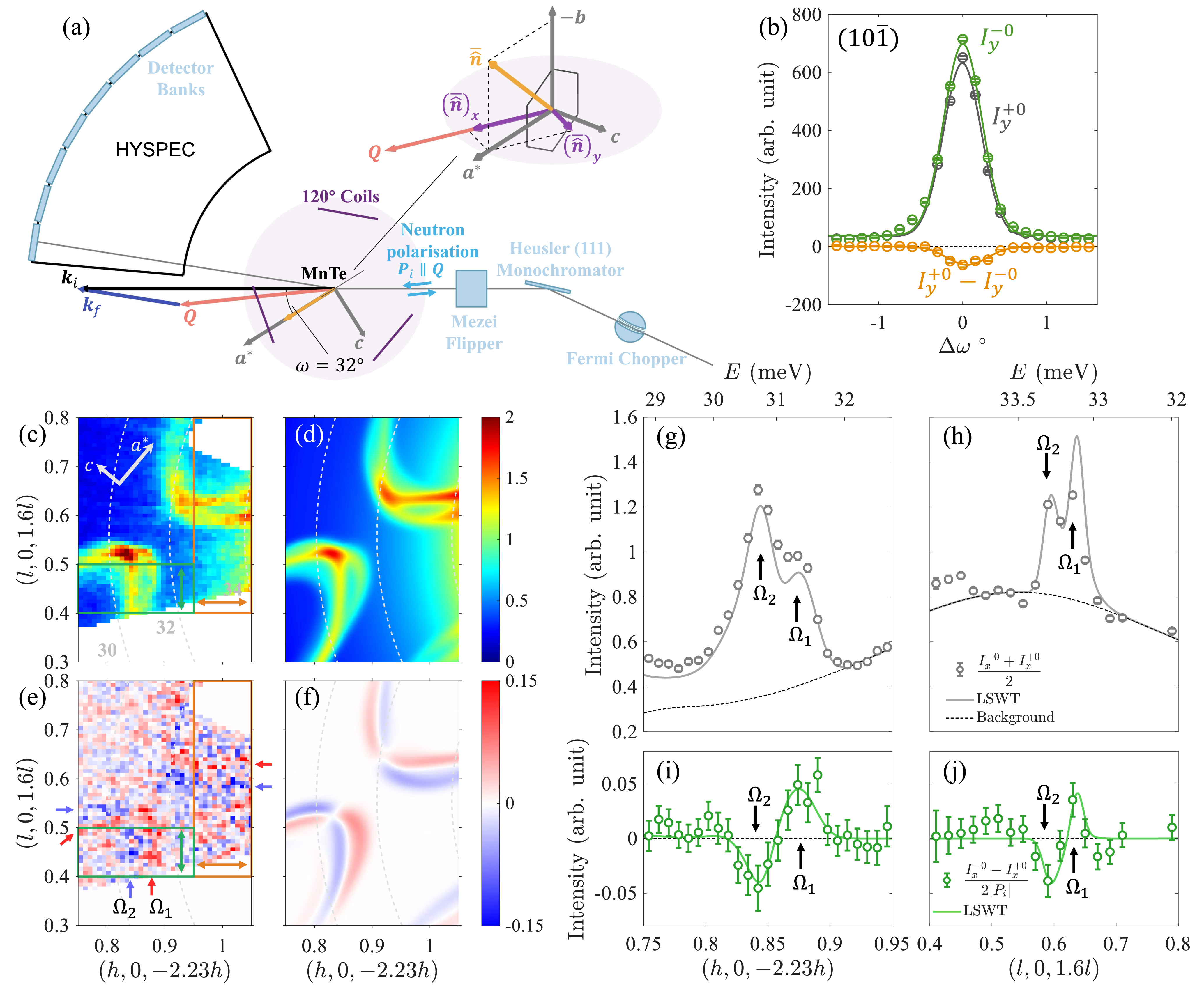}
	\caption{(a) Schematic of half-polarized inelastic neutron scattering at HYSPEC.	During the measurement, the MnTe crystal orientation was fixed at $\omega=32^\circ$.	The inset shows the net unit N\'{e}el vector $\overline{\hat{\boldsymbol{n}}}$ and its components $(\overline{\hat{\boldsymbol{n}}})_x$ and $(\overline{\hat{\boldsymbol{n}}})_y$. (b) Half-polarized neutron diffraction on sample S1 at the Bragg peak $(10\bar{1})$ at 20 K. (c) Symmetric neutron structure factor $\mathbb{S}(\boldsymbol{Q},E)$ measured as the polarization-averaged scattering $(I_x^{-0}+I_x^{+0})/2$ and (d) corresponding LSWT calculation with a background component. Dashed arcs indicate constant-energy contours at $E = 30$, 32, and 34 meV. (e) Chiral term $\mathbb{M}_\text{ch}(\boldsymbol{Q},E)$ measured as $(I_x^{-0}-I_x^{+0})/2|P_{i}|$ and (f) corresponding LSWT calculation. The red and blue arrows mark the magnon modes $\Omega_1$ and $\Omega_2$. The green and orange rectangles in (c),(e) indicate the integration range for the 1D cuts in (g)-(j) along $(h,0,-2.23h)$ and $(l,0,1.6l)$, respectively, with the double arrows indicating the integration range. (g) 1D cut of $(I_x^{-0}+I_x^{+0})/2$ along $(h,0,-2.23h)$, integrated over $(l,0,1.6l)$ from $l=$0.4 to 0.5 and (h) along $(l,0,1.6l)$, integrated over $(h,0,-2.23h)$ from $h = 0.95$ to 1.05. The backgrounds are shown by the dashed line. (i),(j) 1D cut of $(I_x^{-0}-I_x^{+0})/2|P_{i}|$ with the same integration range as (g),(h), respectively.
	}\label{fig2}
\end{figure*}

PINS intensity differences with neutron spin polarized parallel and antiparallel to $\boldsymbol{x}$ probe the chiral term $\mathbb{M}_\text{ch}(\boldsymbol{Q},E) \propto I_{x}^{+-} - I_{x}^{-+}$ (see SM~\cite{Methods} for detail). 
In a collinear altermagnet, $\mathbb{M}_\text{ch}(\boldsymbol{Q},E)$ is proportional to the $x$ component of the net unit N\'{e}el vector, $(\overline{\hat{\boldsymbol{n}}})_x$, as shown in Fig.~\ref{fig2}(a), where the subscript $x$ denotes the projection along the $x$ axis (see SM~\cite{Methods} for detail). 
Thus, observing magnon chirality through $\mathbb{M}_\text{ch}$ requires a nonzero $(\overline{\hat{\boldsymbol{n}}})_x$ in the sample. The IN20 measurements directly provided $I_{x}^{+-} - I_{x}^{-+}$ at each $(\boldsymbol{Q}, E)$. 
HYSPEC measurements, on the other hand, were performed using a half-polarized neutron scattering setup, in which the polarization of the scattered neutrons was not analyzed; nevertheless, the intensity difference between the opposite incident polarization states, $I_{x}^{+0} - I_{x}^{-0}$, remains sensitive to chirality~\cite{maleyevInvestigationSpinChirality1995,maleyevFirstObservationDynamical1998}.

\begin{figure*}[htbp]
	\centering
	\includegraphics[width=\textwidth]{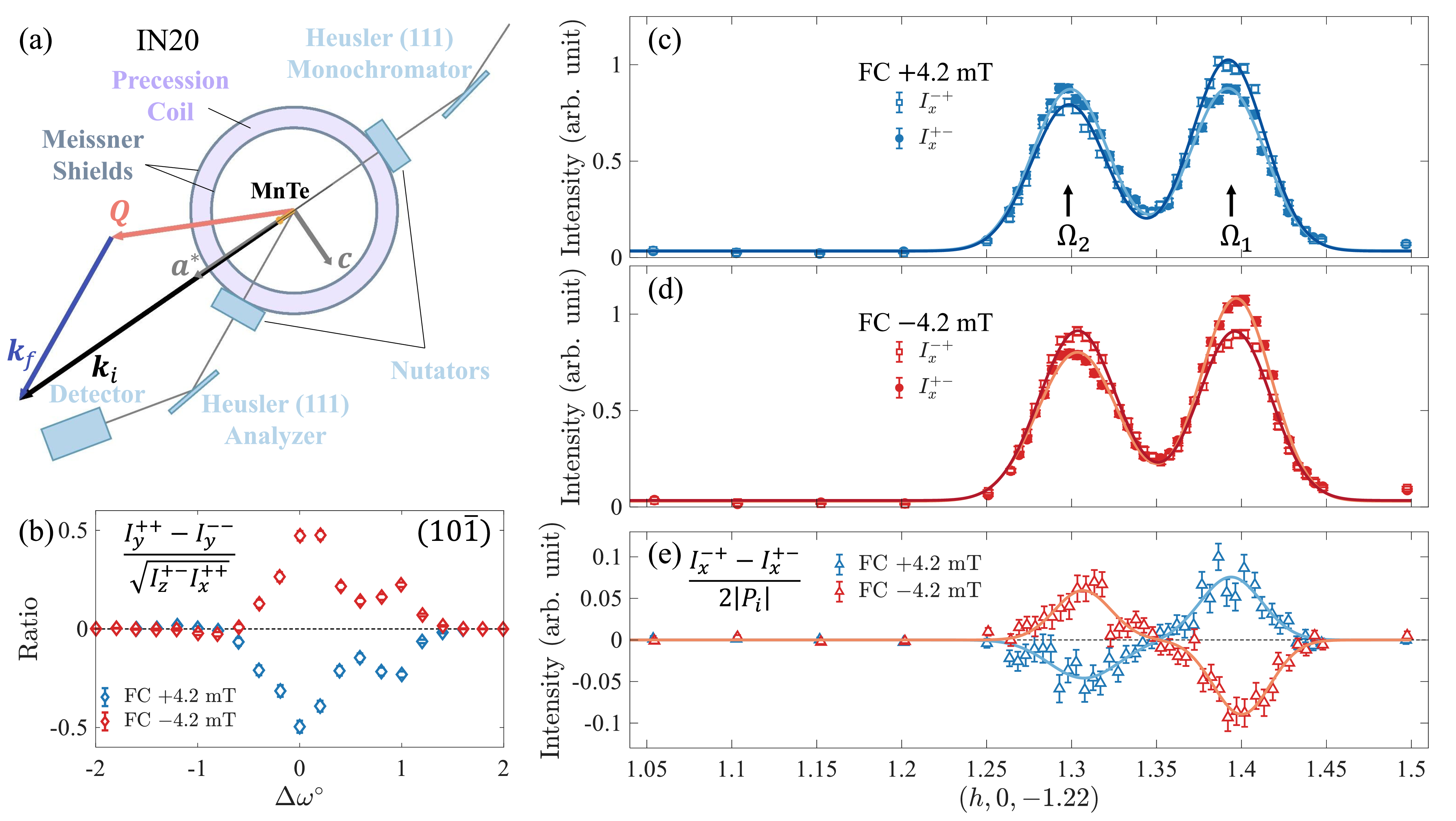}
	\caption{(a) Scheme of the setup of full-PINS at IN20. The incident and scattered neutron spins are manipulated by the nutators and the precession coil. (b) Polarization analysis at the hybrid Bragg peak $(10\bar{1})$ at 10 K. The quantity $(I_y^{++}-I_y^{--})/{\sqrt{I_z^{+-} I_x^{++}}}$ switches sign after FC with $\pm 4.2$~mT. (c)-(e) Constant-$E$ scans at 32 meV of $I_x^{-+}$, $I_x^{+-}$, and $(I_x^{-+}-I_x^{+-})/2|P_i|$ along $(h,0,-1.22)$ after FC with $\pm$4.2 mT. The solid curves show Gaussian fits. The sign reversal of $(I_x^{-+} - I_x^{+-})/(2|P_i|)$ induced by the opposite sign of the cooling field shows the switching of the magnon chirality.
	}\label{fig3}
\end{figure*}

Prior to inelastic spin-wave measurements, a polarized neutron diffraction (PND) measurement of nuclear-magnetic interference (NMI) at the $(10\bar{1})$ Bragg peak was carried out at HYSPEC and IN20 to establish the magnetic-domain imbalance in the sample and thus a non-zero sample-averaged $\overline{\hat{\boldsymbol{n}}}$~\cite{mcclartyObservingAltermagnetismUsing2025}. Polarized incident neutrons with spins along $\boldsymbol{y}$ probe the NMI term, 
$R_y \propto I^{++}_y - I^{--}_y$ (IN20) 
or $R_y \propto I^{+0}_y - I^{-0}_y$ (HYSPEC) (see SM~\cite{Methods} for detail). As shown by the difference between $I^{+0}_y$ and $I^{-0}_y$ in Fig.~\ref{fig2}(b), 
and between $I^{++}_y$ and $I^{--}_y$ in Fig.~\ref{fig3}(b), a nonzero $R_y$ is obtained after FC.  
Because $R_y$ is proportional to $(\overline{\hat{\boldsymbol{n}}})_y$ (see SM~\cite{Methods} for detail), this confirms a finite  $(\overline{\hat{\boldsymbol{n}}})_y$ in the sample. Because the Mn spins lie in the $(hk0)$ plane (see inset of Fig.~\ref{fig2}(a)), a finite $(\overline{\hat{\boldsymbol{n}}})_y$ component measured at a Bragg peak $(h0l)$ with nonzero $l$ implies that $\overline{\hat{\boldsymbol{n}}}$ has a finite projection onto the $(h,0,0)$ axis, which in turn means a nonzero $(\overline{\hat{\boldsymbol{n}}})_x$. Thus, the observation of a finite $R_y$ in PND is a prerequisite for detecting the chiral term $\mathbb{M}_\text{ch}(\boldsymbol{Q},E)$ in the present experimental geometry.

After FC, PINS without polarization analysis after the sample was performed at HYSPEC. 
Here the incident neutron polarization $\boldsymbol{P}_i$ was approximately
parallel to $\hat{\boldsymbol{x}}$, with the sample at 20 K and its orientation fixed to accumulate statistics as illustrated in Fig.~\ref{fig2}(a) (see SM~\cite{Methods} for detail). 
The trajectory of the $\boldsymbol{Q}$-$E$ coverage encounters the magnon band top, where pronounced altermagnetic magnon splitting appears \cite{liuChiralSplitMagnon2024}, as shown in Fig.~\ref{fig1}(c). The splitting is clearly captured by the symmetric term in the neutron structure factor, $\mathbb{S}(\boldsymbol{Q},E)\propto(I_x^{-0}+I_x^{+0})/2$, as shown in Fig.~\ref{fig2}(c). The color plots are displayed using the axes $(h,0,-2.23h)$ and $(l,0,1.6l)$ so that the traces of the magnon-splitting signals appear approximately vertical or horizontal. The one-dimensional (1D) cuts of $\mathbb{S}(\boldsymbol{Q},E)$ along $(h,0,-2.23h)$ and $(l,0,1.6l)$, integrated over the green and orange rectangular regions, respectively, also exhibit split magnon-excitation peaks, as shown in Figs.~\ref{fig2}(g) and~\ref{fig2}(h). The observed magnon excitations are well reproduced by the linear spin-wave theory (LSWT) calculation using the spin Hamiltonian reported in Ref.~\cite{liuChiralSplitMagnon2024}, and including an energy-dependent background, as shown in Figs.~\ref{fig2}(d),~\ref{fig2}(g), and~\ref{fig2}(h) (see SM~\cite{Methods} for detail).

The altermagnetic chiral magnons in MnTe are detected via $\mathbb{M}_\text{ch}(\boldsymbol{Q},E)\propto (I_x^{-0}-I_x^{+0})/2|P_{i}|$, which appears with alternating sign. The 1D cuts in Figs.~\ref{fig2}(i) and~\ref{fig2}(j) clearly display the opposite chirality of the magnon modes $\Omega_1$ and $\Omega_2$ as antisymmetric double peaks, consistent with the LSWT calculation. The chiral magnon signal is also observed in Fig.~\ref{fig2}(e), as indicated by the arrows, and is reproduced by the calculation in Fig.~\ref{fig2}(f). 

We further demonstrate that the magnon chirality can be switched by magnetic-field control. Full polarization analysis was performed at IN20, with incoming and outgoing neutron polarization
adjusted to each $(\boldsymbol{Q},E)$, as shown in Fig.~\ref{fig3}(a).
After FC, 
PND showed a sign reversal of the NMI and thus a
reversal of 
$(\overline{\hat{\boldsymbol{n}}})_y \propto (I_y^{++}-I_y^{--})/\sqrt{I_z^{+-}I_x^{++}}$, as shown in Fig.~\ref{fig3}(b). 
The double-peak feature originates from the mosaicity of the sample S2 and does not affect the constant
energy scans along $(h, 0, -1.22)$ in Figs.~\ref{fig3}(a)-\ref{fig3}(e), where the magnitude of the momentum transfer is varied while the sample orientation changes only minimally.
The nondegenerate magnon modes $\Omega_1$ and $\Omega_2$ are resolved by PINS with $\boldsymbol{P}_i\parallel\hat{\boldsymbol{x}}$ in a constant-$E$ scan at 32 meV, as shown in Figs.~\ref{fig3}(c) and~\ref{fig3}(d). The magnon-excitation peaks are overplotted on the calculated dispersion in Fig.~\ref{fig1}(c), showing excellent agreement. 
After FC with $+4.2$~mT, $\mathbb{M}_\text{ch}(\boldsymbol{Q},E)$ is negative for $\Omega_2$ and positive for $\Omega_1$, as shown in Fig.~\ref{fig3}(e). In contrast, after
FC with $-4.2$ mT, $\mathbb{M}_\text{ch}(\boldsymbol{Q},E)$ is positive for $\Omega_2$ and negative for $\Omega_1$, demonstrating that the sign of the
magnetic field in the FC process controls the reversal of the chirality for both modes.

The ratio $|\mathbb{M}_\text{ch}/\mathbb{S}|$ reflects the magnetic-domain imbalance: it is 0\% for a perfectly balanced multi-domain state and finte for imbalanced multi-domain state (see SM~\cite{Methods} for detail). 
From the HYSPEC PINS data, $|\mathbb{M}_{\mathrm{ch}}/\mathbb{S}|$ is estimated to be 10.5(7)\%. 
The ratio can also be obtained directly from $(\overline{\hat{\boldsymbol{n}}})_x$ determined by PND, yielding 12.2(5)\%, in good agreement with the PINS result.
For the IN20 measurements, $|\mathbb{M}_{\mathrm{ch}}/\mathbb{S}|$ is estimated to be 6.2(4)\% from PINS and 9.7(6)\% from PND, respectively, and these values are mutually consistent. 
This overall consistency demonstrates that the observed chiral-magnon intensities arise exclusively from the imbalanced domain population, thereby confirming the purely altermagnetic origin of magnon chirality in MnTe. Note that though the weak ferromagnetic moment in MnTe~\cite{mazinOriginGossamerFerromagnetism2024,kluczykCoexistenceAnomalousHall2024} also reverses upon switching the sign of the cooling field, its magnitude, on the order of $\sim 10^{-5}\mu_\text{B}$~\cite{kluczykCoexistenceAnomalousHall2024}, is far too small to produce any measurable effect in neutron scattering, thus cannot account for the observed chiral magnons.

We have shown that MnTe, a magnet with zero net magnetization and collinear
order, has magnetic excitations with opposite chirality, a crucial feature of altermagnets. Moreover, the chirality of both magnon modes is controlled by field cooling in a few mT field --- reversing
the sign of the field reverses the chiralities. MnTe possesses a bulk $g$-wave spin symmetry~\cite{smejkalConventionalFerromagnetismAntiferromagnetism2022},  which
prohibits magnon spin-current generation~\cite{wuMagnonSplittingMagnon2025}. Nevertheless, its spin symmetry can be tuned by strain into a
$p$-wave form~\cite{belashchenkoGiantStrainInducedSpin2025}, thereby enabling MnTe to serve as a viable platform for magnonics.
The search for chiral magnons with $d$-wave symmetry in altermagnets~\cite{weiLa$_2$O$_3$Mn$_2$Se$_2$CorrelatedInsulating2025,jiangMetallicRoomtemperatureDwave2025,zhangCrystalsymmetrypairedSpinValley2025} offers another promising direction for applications. Our results establish the first steps toward controlling altermagnetic chiral magnons, which not only advance the emergent field of altermagnetism but also lay the groundwork for its future functional exploitation.

\begin{acknowledgements}
A portion of this research used resources at the Spallation Neutron Source, a DOE Office of Science User Facility operated by the Oak Ridge National Laboratory. The beam time was allocated to HYSPEC (BL-14B) on proposal number IPTS-35462. We greatly appreciate to M. K. Graves-Brook and A. T. Savici for supporting the experiment at HYSPEC. The neutron scattering experiment at IN20 was performed under proposal number 4-01-1895. We are grateful to M.-H. Baurand, Eric Bourgeat-Lami, and Guillaume Bergonzoli for supporting the experiment at ILL. We also acknowledge R. Ishii for assisting with the sample synthesis. Z.L. was supported by the Japan Society for the Promotion of Science through the Leading Graduate Schools (MERIT). This project was supported by JSPS KAKENHI Grant Numbers 21H04441. The work at Brookhaven National Laboratory was supported by the Office of Basic Energy Sciences, U.S. Department of Energy (DOE) under Contract No. DE-SC0012704.
\end{acknowledgements}

{\it Data availability}---The data that support the findings of this article are openly available~\cite{liuSwitchableChiralMagnons2026,masudaObservationAltermagneticChiral2025}.


%

\end{document}